\documentclass[paper,prd,reprint,preprintnumbers,nofootinbib,notitlepage,showpacs,showkeys]{revtex4-2}
\usepackage{latexsym,graphicx,amssymb,amsmath,mathrsfs} 
\usepackage{slashed,mathbbol}
\usepackage[utf8]{inputenc}

\DeclareSymbolFont{bbold}{U}{bbold}{m}{n}
\DeclareSymbolFontAlphabet{\mathbbold}{bbold}

\newcommand{\be}{\begin{equation}}      
\newcommand{\ee}{\end{equation}}      
\newcommand{\bea}{\begin{eqnarray}}      
\newcommand{\eea}{\end{eqnarray}}

\makeatletter
\renewcommand\appendix{\par
\setcounter{section}{0}%   
\setcounter{subsection}{0}% 
\gdef\thesection{\appendixname\space\@Alph\c@section}}

\long\def\unmarkedfootnote#1{{\long\def\@makefntext##1{##1}\footnotetext{#1}}}
\makeatother

\begin{document} 

\title{Electromagnetic energy loss of axion stars} 
\author{A. Patk\'os}
\email{patkos@galaxy.elte.hu}
\affiliation{Institute of Physics, E\"otv\"os University, 1117 P\'azm\'any P\'eter s\'et\'any 1/A, Budapest, Hungary}

\begin{abstract}
The rate of energy exchange of spatially localized and gravitationally bound axion configurations with electromagnetic radiation is investigated in presence of strong static magnetic fields. 
Fully analytic treatment is achieved based on variationally optimized separable spatio-temporal clump profiles. For dilute axion stars the equation of the energy variation is reinterpreted as a rate equation of the axion number. Its solutions hints to an asymptotic $\sim t^{1/5}$ increase of the clump size. 
\end{abstract}

\maketitle

\section{Introduction}
A boson star is expected to form of free massive scalar particles merely due to their gravitational interaction. Such objects were studied first numerically by Kaup \cite{kaup68}, and Ruffini and Bonnazola \cite{ruffini69}. In the context of axions, a leading  dark matter candidate, this phenomenon was first investigated by Tkachev \cite{tkachev86}.  The characteristics of axionic scalar stars has been extensively explored in the literature \cite{tkachev91, kolb93,kolb94,chavanis11a,chavanis11b}. Numerical studies of Chavanis and Delfini \cite{chavanis11} including also quartic self-interaction of axions pointed out the existence of a stability edge in the mass-radius relation for this so-called {\it dilute} branch. Semi-analytical solutions in the gravity dominated branch were constructed by Eby {\it al.}\cite{eby15}. Another branch where higher n-point axion interactions can stabilize even high mass localized configurations,  were discovered by Braaten {\it et al.}\cite{braaten16} and called {\it dense} axion stars.

In the equations of gravitationally bound axion stars the energy is dominated by the rest-mass of axions and one can apply the non-relativistic approximation. Then the axion number content of a field configuration is an approximately conserved quantity. When one includes via higher order couplings the generation of more energetic (relativistic) particles then some particles get the chance to escape from the Bose-Einstein condensate of the star. This effect can be accounted for by introducing a non-Hermitian term into the axion potential \cite{braaten17} which arises from scattering processes of the full theory involving the  high momentum tail of the axion field \cite{eby16a,eby16b}. In case of the dilute branch the main mechanism leading to the depletion of the axionic medium is the two-photon decay of axions\cite{sikivie83}.  A detailed discussion of the lifetime of axion stars is found in the review of Braaten and Zhang \cite{braaten19}.

Strong external magnetic fields are present around neutron stars ($10^4 - 10^{11}$)T, in particular around magnetars ($10^9 - 10^{11}$)T. Electromagnetic radiation from axion stars embedded in strong magnetic fields has been studied in Refs.\cite{amin21, sen22a}.  A generic dimensionless combination of some axion data and the magnetic field strength was suggested to determine the order of magnitude of the electromagnetic decay time in Ref.\cite{arvinataki10}. A computational algorithm to include the radiation back reaction effect quantitatively into the evolution of dense branch stars was put forward and implemented very recently\cite{sen22}.

The aim of the present short paper is to outline a fully analytic computation of the time evolution of the particle number of gravitationally bound (dilute) axion clump in strong magnetic field. The treatment relies on the discussion of the energy-momentum transfer between axions and the electromagnetic field, analyzed in our previous publication\cite{patkos22}. The rate equation for the change of the energy in the axion sector will be analyzed, taking into account fully the back reaction of the emerging electromagnetic radiation (section 2). In the course of the calculation a spherically symmetric axion configuration is assumed. Its separable spatio-temporal profile function is given in terms of variationally optimized trial functions. Their variational treatment is reviewed in section 3. Based on such profile functions, a rather compact analytic formula is given in section 4 for the right hand side of the energy rate equation. It is translated into the rate equation of the axion number in section 5, which is solved explicitly for two trial spatial profiles used already in the previous literature. The asymptotic solution of this equation leads to conjecturing {\it powerlike} time dependence for the particle depletion of axion stars due to electromagnetic radiation. 

\section{Electromagnetic energy balance of an axion clump}

Temporal and spatial variation of axion field $a({\bf x},t)$ in presence of a static magnetic background field $B_0({\bf x})$ represent effectively electromagnetic source densities, the strength of which is determined by the axion--two-photon coupling $g_{a\gamma\gamma}$ \cite{sikivie83,sikivie21}:
\begin{eqnarray}
&\displaystyle
{\bf j}_a({\bf x},t)=-g_{a\gamma\gamma}\dot a({\bf x},t){\bf B}_0({\bf x}),\nonumber\\
&\displaystyle
 \rho_a({\bf x},t)=g_{a\gamma\gamma}{\bf B}_0({\bf x})\cdot\nabla a({\bf x},t).
\label{axion-em-density}
\end{eqnarray}
This means that in Lorentz-gauge one finds the following retarded scalar and vector potentials generated by the axions, using natural units ($\hbar=c=1$) :
\begin{eqnarray}
&\displaystyle
{\bf A}({\bf x},t)=
\int d^3x^\prime\frac{{\bf j}_a({\bf x}^\prime,t-|{\bf x}-{\bf x}^\prime|)}{|{\bf x}-{\bf x}^\prime|},\nonumber\\
&\displaystyle
A_0({\bf x},t)=
\int d^3x^\prime\frac{\rho_a({\bf x}^\prime ,t-|{\bf x}-{\bf x}^\prime|)}{|{\bf x}-{\bf x}^\prime|}).
\label{potential-expressions}
\end{eqnarray}
It is worthwhile to emphasize that $|{\bf B}_0({\bf x})|>>|{\bf\nabla}\times{\bf A}({\bf x},t)|$, by assumption, throughout the investigation below.
The rate of  the energy exchange of any axion configuration with electromagnetic fields formally coincides with the expression of the work power of electrically charged currents, as it was emphasized recently by various authors \cite{sen22,patkos22}, (see also the Appendix of Ref.\cite{paixao22}):
\begin{eqnarray}
&\displaystyle
\frac{dE_a}{dt}=\int d^3x{\bf j}_a({\bf x},t)\cdot{\bf E}({\bf x},t)\nonumber\\
&\displaystyle
=\int d^3xg_{a\gamma\gamma}\dot a({\bf x},t){\bf B}_0({\bf x})\cdot\left(\dot{\bf A}({\bf x},t)+\nabla_{\bf x}A_0({\bf x},t)\right).
\label{energy-work-rate}
\end{eqnarray}
One might note also that the rate of momentum change of the axion clump due to the presence of topologically non-trivial electromagnetic field density is given as
\begin{eqnarray}
&\displaystyle
\frac{d{\bf P}_a}{dt}=\int d^3xg_{a\gamma\gamma}\nabla a({\bf x},t)\nonumber\\
&\displaystyle
\times\left[{\bf B}_0({\bf x})\cdot\left(-\dot{\bf A}({\bf x},t)-\nabla_{\bf x}A_0({\bf x},t)\right)\right].
\label{em-momentum-transfer}
\end{eqnarray}
One substitutes the expressions (\ref{potential-expressions}) of the potentials into (\ref{energy-work-rate}) and performs in the second term (involving the scalar potential) a partial integration. Then one can exploit the continuity equation $\partial_t\rho_a=-\nabla {\bf j}_a$, valid also for the axionic "charge" and "current" densities to arrive at 
\begin{eqnarray}
&\displaystyle
\frac{dE_a}{dt}=\int d^3x\int d^3x^\prime
\frac{1}{|{\bf x}-{\bf x}^\prime|}\nonumber\\
&\displaystyle
\times\Bigl[{\bf j}_a({\bf x},t)\cdot\frac{\partial}{\partial t}{\bf j}_a({\bf x}^\prime,t-|{\bf x}-{\bf x}^\prime|)\nonumber\\
&\displaystyle
+\rho_a({\bf x}^\prime,t-|{\bf x}-{\bf x}^\prime|)\frac{\partial}{\partial t}\rho_a({\bf x},t)\Bigr].
\label{general-axion-energy-balance}
\end{eqnarray}
The clump oscillates with some average frequency $\omega_a$. The retardation dependence of  $\rho_a, {\bf j}_a$ can be expanded into Taylor series in the near-zone, where $\omega_a|{\bf x}-{\bf y}|<<1$. Separating the first term of the expansions from the rest, one finds the time derivative of an "electromagnetic" contribution to the energy of the axion configuraton, $\Delta E_a$. It has the obvious interpretation being the electrostatic and magnetostatic energy of the near zone promptly following the oscillation of the corresponding source densities.  The rest ($W_{rad-loss}$) can be associated with the energy lost by the clump via electromagnetic radiation.
\begin{eqnarray}
&\displaystyle
\frac{dE_a}{dt}
=-\frac{d\Delta E_a}{dt}-W_{rad-loss}\nonumber\\
&\displaystyle
\Delta E_a=\frac{1}{2}g_{a\gamma\gamma}^2\int d^3x\int d^3x^\prime\frac{B_{0i}({\bf x})B_{0j}({\bf x}^\prime)}{|{\bf x}-{\bf x}^\prime|}
\nonumber\\
&\displaystyle
\times\left[\delta_{ij}\dot a({\bf x},t)\dot a({\bf x}^\prime,t)+\nabla_{ x_i}a({\bf x},t)\nabla_{x^\prime_j}a({\bf x}^\prime,t)\right].
\label{kieg-axion-energia}
\end{eqnarray}
One might attempt to represent the energy loss with just the next term of the Taylor expansion. Also one notes (exploiting the continuity equation) that there is no contribution from the second ("electric") term proportional to $(\int d^3x\dot\rho({\bf x},t))^2$. The "magnetic" piece reveals suggestive analogy with the so-called  "three-dot" force exerted by the emitted electromagnetic radiation on the motion of its pointlike electric source: 
\begin{eqnarray}
&\displaystyle
W_{rad-loss}\nonumber\\
&\displaystyle
=-g_{a\gamma\gamma}^2\int d^3x \dot a({\bf x},t){\bf B}_0({\bf x})\cdot\int d^3x^\prime \dddot a({\bf x}^\prime,t){\bf B}_0({\bf x}^\prime).
\label{veszteseg-three-dots}
\end{eqnarray}
One can average this expression over the average period $T=2\pi/\omega_a$ and obtain with partial time integration the compact formula
\begin{equation}
\overline{W_{rad-loss}}^T=g_{a\gamma\gamma}^2\overline{\left(\int d^3x\ddot a({\bf x},t){\bf B}_0({\bf x})\right)^2}^T.
\label{veszteseg-three-dots-averaged}
\end{equation}

 In the next section we summarize some well-established facts about axion stars of spherically symmetric shape also assuming separable time and space dependence of their profile. A simple variational approximation to its binding energy allows the determination of its size and establishing a functional relation with the number of its constituting axions. It becomes clear that for gravitationally bound axion clumps one cannot justify the above Taylor-expansion, one has to find an exact treatment for retardation effects.

\section{Profile of an axion star}

 Here we outline the construction steps and the main features characterising the spatial profile of scalar stars emerging from an equilibrium between gravitational attraction and kinetic pressure. For this a reduction to a non-relativistic approximation of the full theory is performed. Mostly, we follow the treatments of Refs. \cite{guth15,eby16b}. 

The Hamiltonian governing the dynamics of the axion condensate in its own gravitational field reads: 
\begin{eqnarray}
&\displaystyle
H=\int d^3x\left[\frac{1}{2}\left(\dot a({\bf x},t)\right)^2+\frac{1}{2}\left(\nabla a({\bf x},t)\right)^2+\frac{1}{2}m_a^2a({\bf x},t)^2\right]
\nonumber\\
&\displaystyle
+U_{grav}~~~~~~~~~~~~~~~~~~~~~~~~~~~~~~~~~~~~~~~~.
\label{clump-hamiltonian}
\end{eqnarray}
The gravitational energy is determined by the mass-density distribution $\rho_{mass}({\bf x},t)$ of axions:
\begin{equation}
U_{grav}=-\frac{G_N}{2}\int d^3x\int d^3x^\prime\frac{\rho_{mass}({\bf x},t)\rho_{mass}({\bf x}^\prime,t)}{|{\bf x}-{\bf x}^\prime|}.
\label{gravitational-energy}
\end{equation}
In the condensate all particles have nearly the rest mass energy $m_a$, therefore the field $a({\bf x},t)$ is parametrized as the product  of the corresponding harmonic oscillation with a slowly varying amplitude. 
\begin{eqnarray}
&\displaystyle
a({\bf x},t)=\frac{1}{\sqrt{2m_a}}\left(e^{-im_at}\psi({\bf x},t)+e^{im_at}\psi^*({\bf x},t)\right),\nonumber\\
&\displaystyle
 \psi({\bf x},t)=e^{-i\mu_gt}\tilde\psi({\bf x}).
\label{a-ansatz}
\end{eqnarray}
The time dependence of the slowly varying term is approximated by a small frequency shift $\mu_g<<m_a$. When substituting this Ansatz into the canonical equations one keeps only the first time-derivative of $\psi({\bf x},t)$ in view of its assumed slow variation. This results in the following equation: 
\begin{equation}
i\dot\psi=-\frac{1}{2m_a}\bigtriangleup\psi+\frac{\delta U_{grav}}{\delta(\psi^*\psi)}\psi.
\label{schroedinger-like-equation}
\end{equation}
As a consequence one can define a density:  $\psi^*\psi$, which is conserved in the present approximation. Its integral is identified with the axion number $N$ of the clump.
\begin{equation}
\frac{d}{dt}(\psi^*\psi)=0,\rightarrow N_a=\int d^3x |\tilde\psi({\bf x})|^2={\textrm{const.}}.
\label{normalisation}
\end{equation}
The corresponding mass density $\rho_{mass}=m_a\tilde\psi^*\tilde\psi$ can be used in the expression of the gravitational energy:
\begin{equation}
 \qquad U_{grav}=-\frac{G_N}{2}\int d^3x\int d^3x^\prime\frac{(m_a|\tilde\psi({\bf x})|^2)( m_a|\tilde\psi({\bf x^\prime})|^2)}{|{\bf x}-{\bf x}^\prime|}.
\label{gravitational-energy--with-ansatz}
\end{equation}
Substituing  (\ref{a-ansatz}) into the equation of $\psi$ one arrives at an eigenvalue equation for $\mu_g$.
\begin{equation}
\mu_g\tilde\psi({\bf x})=-\frac{1}{2m_a}\bigtriangleup\tilde\psi({\bf x})+\tilde\psi({\bf x})
G_N\int d^3x^\prime\frac{ m_a^2|\tilde\psi({\bf x^\prime})|^2}{|{\bf x}-{\bf x}^\prime|}.
\label{eigenvalue-frequency-shift}
\end{equation}
The corresponding non-relativistic Hamilton-operator is readily identified and its minimum in the space of $\tilde\psi$ functions determines the binding energy of the clump:
\begin{eqnarray}
&\displaystyle
H_{non-rel}=\int d^3x\frac{1}{2m_a}|\nabla\tilde\psi|^2+U_{grav},  \nonumber\\
&\displaystyle
 E_{ground}=min[H_{non-rel}(\tilde\psi)].
\label{non-rel-hamiltonian}
\end{eqnarray}
The ground state energy and the gravitatonal energy of the ground state profile determine the frequency shift due to the binding:
\begin{equation}
N\mu_g=\int d^3x\frac{1}{2m_a}|\nabla\tilde\psi|^2+2U_{grav}=E_{ground}+U_{grav}.
\label{shift-determination}
\end{equation}

In the present note we will be satisfied with a variational estimate after choosing a spherically symmetric ansatz for $\tilde\psi$:
\begin{equation}
\tilde\psi({\bf x})=wF(\xi),\quad \xi=\frac{|{\bf x}|}{R}, \qquad w^2=\frac{N}{C_2R^3}.
\label{scaled-profile}
\end{equation}
Here we have introduced the parameter $R$ to be determined variationally, which characterizes the spatial extension of the clump. The quantity $w$ is determined by the normalisation condition
(\ref{normalisation}). The energy function in units of $m_a$ depends on the dimensionless parameters ($m_aR,N$) and also on the combination $G_Nm_a^2$ of the physical constants: 
\begin{equation}
E(R,N)=m_a\left(\frac{D_2}{2C_2}\frac{1}{(m_aR)^2}-\frac{B_4}{2C_2^2}\frac{1}{m_aR}G_Nm_a^2N\right),
\label{two-parameter-energy-function}
\end{equation}
where specific integrals over the profile function $F(\xi)$ are introduced:
\begin{eqnarray}
&\displaystyle
C_2=4\pi\int_0^\infty d\xi \xi^2 F^2(\xi),\qquad D_2=4\pi\int_0^\infty d\xi \xi^2 F^{\prime 2}(\xi),\nonumber\\
&\displaystyle
B_4=32\pi^2\int_0^\infty d_\xi F^2(\xi)\int_0^\xi d\eta\eta^2F^2(\eta).
\label{profile-integrals}
\end{eqnarray}
One minimizes $E(R,N)$ at fixed value of $N$ with respect of $m_aR$ which yields:
\begin{eqnarray}
&\displaystyle
X\equiv (m_aR)_{opt}=\frac{2C_2D_2}{B_4}\frac{1}{G_Nm_a^2N},\nonumber\\
&\displaystyle
 E_a=Nm_a+E_{ground},
 \qquad N\mu_g=\frac{3}{2}E_{ground},\\
&\displaystyle
E_{ground}=-m_a\frac{B_4^2}{8C_2^3D_2}(G_Nm_a^2N)^2\equiv E_g(X(NG_nm_a^2)).\nonumber
\label{variational-parametrisation}
\end{eqnarray}
The order of magnitude of the coefficient built from integrals over the clump profile is $\le {\cal O}(10^2)$. Choosing for the axion mass $m_a=10^{-14}$GeV, one finds $G_Nm_a^2\approx 5\times 10^{-66}$. Guth {\it et al.} \cite{guth15} argue that at the formation of axion stars (near the temperature of the QCD transition) $N\approx 10^{61}$. Using these values one finds that $X>>1$ and $\mu_g<<m_a$. Retardation effects can not be treated perturbatively for this solution.

The full energy of the axionic clump is the sum of the rest-masses of the free axions and the binding energy. Both $X$ and $E_g$ depend parametrically on $G_Nm_a^2N\sim 5\cdot 10^{-66}N$.

Since one can neglect here the frequency shift $\mu_g$ relative to $m_a$, choosing real functions for $F(\xi)$ we use below the following trial function with separable time and radial dependence:
\begin{equation}
a_{dilute}=\sqrt{\frac{2}{m_a}}\cos(m_at)wF(\xi).
\label{no-shift-a-ansatz}
\end{equation}

\section{Electromagnetic energy loss including retardation effects}

We return to the evaluation of the rate of energy change of the axion clump (\ref{general-axion-energy-balance}) after substituting (\ref{no-shift-a-ansatz}) into (\ref{axion-em-density}). In the "magnetic" and the "electric" parts of the integrand the following expressions multiply the squared photon-axion coupling:
\begin{eqnarray}
&\displaystyle
\dot a({\bf x},t)\ddot a({\bf x}^\prime,t-|{\bf x}-{\bf x}^\prime|)\nonumber\\
&\displaystyle
=2m_a^2\sin(m_at)\cos\left(m_a(t-|{\bf x}-{\bf x}^\prime|)\right)w^2F(\xi_x)F(\xi_{x^\prime}),\nonumber\\
&\displaystyle
\left({\bf B}_0\cdot\nabla_x\dot a({\bf x},t)\right)\left( {\bf B}_0\cdot\nabla_{x^\prime} a({\bf x}^\prime,t-|{\bf x}-{\bf x}^\prime|)\right)\nonumber\\
&\displaystyle
=2B_0^2\left({\bf n}_B\cdot\frac{\hat{\bf x}}{R}\right)\left({\bf n}_B\cdot\frac{\hat{\bf x}^\prime}{R}\right)\nonumber\\
&\displaystyle
\times\sin(m_at)\cos\left(m_a(t-|{\bf x}-{\bf x}^\prime|)\right)w^2F^\prime(\xi_x)F^\prime(\xi_{x^\prime}).
\label{explicit-intagrands}
\end{eqnarray}
In the third line the gradient operator applies only to the first arguments.
The time average over the period $T=2\pi/m_a$  is easily performed with help of simple trigonometric identities leading to
\begin{eqnarray}
&\displaystyle
\overline{\frac{dE_a}{dt}}^T
=-\int d^3x\int d^3x^\prime \frac{g_{a\gamma\gamma}^2B_0^2w^2}{|{\bf x}-{\bf x}^\prime|}\sin\left(m_a|{\bf x}-{\bf x}^\prime|\right)\\
&\displaystyle
\times\left[m_a^2F(\xi_x)F(\xi_{x^\prime})+
\left({\bf n}_B\cdot\frac{\hat{\bf x}}{R}\right)\left({\bf n}_B\cdot\frac{\hat{\bf x}^\prime}{R}\right)F^\prime(\xi_x)F^\prime(\xi_{x^\prime})\right].\nonumber
\label{egzakt-veszteseg}
\end{eqnarray}
One rediscovers the "three-dot" result for the energy loss (\ref{veszteseg-three-dots-averaged}) when one keeps only the first term of Taylor-series of $\sin(m_a|{\bf x}-{\bf x}^\prime|)$. The near-zone electromagnetic contribution to the energy of the axion clump (\ref{kieg-axion-energia}) oscillates periodically, therefore its average vanishes. Now, one proceeds with the evaluation of the angular parts of the space integrations by exploiting the following factorisation of the $|{\bf x}-{\bf x}^\prime|$ dependence of the integrand:
\begin{eqnarray}
&\displaystyle
%{\textrm{Im}}
\frac{e^{im_a|{\bf x}-{\bf x}^\prime|}}{4\pi|{\bf x}-{\bf x}^\prime|}\\
\label{factorisation}
&\displaystyle
=
%{\textrm{Im}}
\left[im_a\sum_{l=0}^\infty j_l(m_ar_<)h_l^{(1)}(m_ar_>)\sum_{m=-l}^lY_{lm}^*(\hat{\bf x})Y_{lm}(\hat{\bf x}^\prime)\right],\nonumber
\end{eqnarray}
where $r_<={\textrm{min}}(|{\bf x}|,|{\bf x}|^\prime), r_>={\textrm{max}}(|{\bf x}|,|{\bf x}|^\prime)$. In the actual expression of the integrand of the time averaged energy loss one has to take the imaginary part of both sides.
The angular integrations of the "magnetic" term receive contribution only from the $l=0$ spherical harmonics, while in the "electric" term only $Y_{10}$  contributes when the  $z$-axis is chosen along $\hat{\bf n}_B$. The relevant imaginary parts turn out invariant for the exchange of $r_<$ and $r_>$:
\begin{equation}
{\textrm{Im}}\left[ij_0(m_ar_<)h_0^{(1)}(m_ar_>)\right]=\frac{\sin(m_ar_<)}{m_ar_<}\cdot\frac{\sin(m_ar_>}{m_ar_>}.
\label{imaginary-part-0}
\end{equation}
\begin{eqnarray}
&\displaystyle
{\textrm{Im}}[ij_1(m_ar_<)h_1^{(1)}(m_ar_>)]
=\left(\frac{\sin(m_ar_<)}{(m_ar_<)^2}-\frac{\cos(m_ar_<)}{m_ar_<}\right)\nonumber\\
&\displaystyle
\times
\left(\frac{\sin(m_ar_>)}{(m_ar_>)^2}-\frac{\cos(m_ar_>)}{m_ar_>}\right).
\label{imaginary-part-1}
\end{eqnarray}
The $r_<\leftrightarrow r_>$ invariance has the consequence that the radial $\xi_x$ and $\xi_{x^\prime}$ integrations are independent and lead to the same result. By this observation one finds a rather compact expression for the averaged electromagnetic energy loss of an axion star:
\begin{eqnarray}
&\displaystyle
\overline{\frac{dE_a}{dt}}^T=-m_a\cdot m_aN\cdot \frac{g_{a\gamma\gamma}^2B_0^2}{m_a^2}\frac{X^3}{C_2}\left(I^2_{mag}+\frac{1}{3X^2}I^2_{el}\right)\nonumber\\
&\displaystyle
\equiv
-m_a\cdot m_aN\cdot \frac{g_{a\gamma\gamma}^2B_0^2}{m_a^2}{\cal F}(X(G_Nm_a^2N)) ,\nonumber\\
&\displaystyle
I_{mag}=\int d^3\xi\frac{\sin(X\xi)}{X\xi}F(\xi),\nonumber\\
&\displaystyle
 I_{el}=\int d^3\xi\left(\frac{\sin(X\xi)}{(X\xi)^2}-\frac{\cos(X\xi)}{X\xi}\right)F^\prime(\xi).
\label{exact-analytic-energy-loss}
\end{eqnarray}
The first three terms on the right hand side follow the parametric form denoted as $\tau_\gamma$ in Ref.\cite{arvinataki10}. The temporal dependence of the decay however is essentially influenced by the $X$-dependent (in the final count $N$-dependent) factor ${\cal F}(X)$. This effect has been observed in Refs.\cite{sen22a,amin21}.  In case of clumps stabilized by the self-interaction of axions Refs.\cite{amin21,sen22} have found numerically ${\cal F}\approx 2$. These investigations all arrive at the conclusion that axionic clumps decay exponentially. The only difference is in the scaled value of the  decay constant $m_a\tau_\gamma$. With help of specific choices of $F(\xi)$ we shall evaluate this factor and establish its limiting behavior in the extreme dilute ($X>>1$) case, relevant to the above chosen values of the physically free parameters. It will be shown that this leads to a qualitatively different conclusion concerning the temporal dependence of the number of particles in an axion star, $N(t)$.

It is notable that a completely analogous analysis of the rate of momentum transfer given in (\ref{em-momentum-transfer}) arising from the magnetically induced electromagnetic radiation to the spherically symmetric axion clump leads to the conclusion that the resulting force is zero.

\section{Time evolution of axion particle number for generic clump profiles}

Several specific trial profile functions were used for estimating the energy of axionic clumps. Guth {\it et al.}\cite{guth15} have chosen a simple exponential by the example of the ground state wave function of the hydrogen atom, although it does not satisfy the prescribed boundary condition $\lim_{r\rightarrow 0}\partial_r\tilde\psi (|{\bf x}|)=0$.  In Ref.\cite{eby16b} a Gaussian was used for the variational estimation. For our analytic calculation an alternative cosine profile is more convenient, which was shown \cite{eby16b} to reproduce the Gaussian very well in the range $|{\bf x}|<R$. We have computed the profile integrals (\ref{profile-integrals}) and the factor $\cal F$ (\ref{exact-analytic-energy-loss}) with the following trial profiles:
\begin{eqnarray}
&\displaystyle
F^{exp}=e^{-\xi},\qquad X^{exp}\approx 20(G_Nm_a^2N)^{-1},\nonumber\\
&\displaystyle
 E_g^{exp}\approx -0.31(G_Nm_a^2N)^2m_a,
\end{eqnarray}
and
\begin{eqnarray}
&\displaystyle
F^{cos}=\cos^2\left(\frac{\pi \xi}{2}\right),~~ \xi<1,\quad X^{cos}\approx11.2(G_Nm_a^2N)^{-1}, \nonumber\\
&\displaystyle
E_g^{cos}\approx -0.25(G_Nm_a^2N)^2m_a.
\end{eqnarray}
The elementary integrals which determine $I_{mag}, I_{el}$ in (\ref{exact-analytic-energy-loss}) lead to apparently quite different expressions for the average energy loss per particle mass:
 \begin{equation}
\frac{1}{m_aN}\overline{\frac{dE_a^{exp}}{dt}}^T=-m_a\left(\frac{g_{a\gamma\gamma}B_0}{m_a}\right)^2\frac{256\pi}{3}\frac{X^3}{(1+X^2)^4},
\label{rate-exp}
\end{equation}
\begin{eqnarray}
&\displaystyle
\frac{1}{m_aN}\overline{\frac{dE_a^{cos}}{dt}}^T=-m_a\left(\frac{g_{a\gamma\gamma}B_0}{m_a}\right)^2\frac{4\pi^2}{C_2}\nonumber\\
&\displaystyle
\times\Biggl\{X\left[\frac{\cos X}{X}\frac{\pi^2}{X^2-\pi^2}+\sin X\left(\frac{1}{X^2}-\frac{X^2+\pi^2}{(X^2-\pi^2)^2}\right)\right]^2\nonumber\\
&\displaystyle
+\frac{\pi^4}{3 X(X^2-\pi^2)^2}\left[\cos X-\frac{\sin X }{X}\frac{3X^2-\pi^2}{X^2-\pi^2}\right]^2\Biggr\}.
\label{rate-cos}
\end{eqnarray}
However, the asymptotic behaviors both for small and very large values of $X$ turn out the same:
\begin{equation}
{\cal F}(X(G_Nm_a^2N))\sim X^3,~~~X<<1, ~~~~~\sim X^{-5},~~~X>>1.
\end{equation}
The coinciding asymptotic behaviours prompt the conjecture that they reflect the nature of the exact solution. 

This conjecture can be actually proven for profile functions which are nonzero in the interval $\xi\in (0,\Lambda)$ and fulfill at the upper end the boundary conditions
\be
F(\Lambda)=F^\prime(\Lambda)=0.
\label{boundary-cond}
\ee
The asymptotic behavor can be constructed in the same steps both for $I_{magn}$ and $I_{el}$. Here we give some details for the first one.  The limiting behavior when $X\rightarrow 0$ is easily established by performing the limit directly in the integrand:
\be
\lim_{X\rightarrow 0}I_{magn}=\int d^3\xi F(\xi).
\ee
For the large  $X$ asymptotics it is convenient to introduce the integration variable $u=X\xi$ and perform three(!) partial $u$-integrations, taking into account the boundary conditions (\ref{boundary-cond}) in the intermediate steps. This results in
\bea
&\displaystyle
\lim_{X\rightarrow \infty}I_{magn}\nonumber\\
&\displaystyle
=-\frac{8\pi}{X^4}F^\prime(0)+\frac{4\pi}{X^5}\left(\Lambda X\cos(\Lambda X)+3\sin(\Lambda X)\right)F^{\prime\prime}(\Lambda)\nonumber\\
&\displaystyle
+\frac{4\pi}{X^6}\int_0^{\Lambda X}du (3\sin u-u\cos u)F^{\prime\prime\prime}(u/X)
\eea
One promptly recognizes that the contribution from the last integral is at most ${\cal O}(X^{-5})$, therefore the leading asymptotics reads as
\be
\lim_{X\rightarrow \infty}I_{magn}
=-\frac{8\pi}{X^4}F^\prime(0)+\frac{4\pi}{X^4}\Lambda \cos(\Lambda X)F^{\prime\prime}(\Lambda)
\ee
In case of the exponential ansatz one has $F^{exp,\prime}(0)=-1, F^{exp,\prime\prime}(\Lambda)=0$, for the cosine ansatz $F^{cos,\prime}(0)=0, F^{cos,\prime\prime}(\Lambda)=\pi^2/2$
and the above formula reproduces the asymptotics directly obtained from (\ref{rate-exp}) and (\ref{rate-cos}).  

The same steps lead for $I_{el}/X^2$ to the same $\sim X^{-4}$ behavior in the above class of axion clump profiles. The class is generic enough (although not being the most general) to strengthen conjecturing this asymptotics also for the exact  solution. Now we turn to the rate equation of the axion number.

In regions where the gravitational binding energy is small, one can use the approximate relation $E_a\approx Nm_a$ which allows to reinterpret the rate equation of the energy as an equation for the rate of change of the axion number of the clump: 
\begin{equation}
\frac{1}{m_a}\frac{1}{m_aN}\overline{\frac{dE_a}{dt}}^T\approx\frac{1}{N} \frac{dN}{d(m_at)}= -\frac{g_{a\gamma\gamma}^2B_0^2}{m_a^2}{\cal F}(X(G_Nm_a^2N)).
\end{equation}
This can be certainly applied for $X>>1$ and eventually one finds a slow algebraic blowing up of the clump which is accompanied (in view of  the inverse relation of $X $ and $N$) by the diminishing of the number of axions due to electromagnetic radiation:
\begin{equation}
X(t)\sim (G_Ng_{a\gamma\gamma}^2B_0^2m_at)^{1/5}, \qquad X>>1.
\end{equation}

\section{Summary}
Starting from the equation gonverning the energy balance of the axion sector of axion electrodynamics a dynamical rate equation has been derived for the particle number content of a gravitationally bound axion clump. Our fully analytic treatment was based on assuming a separable single frequency, spherically symmetric ansatz for the spatio-temporal profile of the axion star. Variational treatment of two rather different looking explicit trial spatial profile functions led to the same behavior of relative axion number rate both in the small-size and large-size regimes. This experience led us to conjecture a universal asymptotic form of the rate equation for large axion numbers. In this region the characteristic size parameter of the object slowly increases as $\sim t^{1/5}$.

\end{document}